# Quantum Interferences of Pseudospin-mediated Atomic-scale Vortices in Monolayer Graphene

Yu Zhang, Lin He[†]

Center for Advanced Quantum Studies, Department of Physics, Beijing Normal University, Beijing, 100875, People's Republic of China

[†]Correspondence and requests for materials should be addressed to Lin He (e-mail: helin@bnu.edu.cn).

**Vortex is a universal and significant phenomenon that has been known for centuries. However, creating vortices to the atomic limit has remained elusive because that the characteristic length to support a vortex is usually much larger than the atomic scale. Very recently, it was demonstrated that intervalley scattering induced by the single carbon defect of graphene leads to phase winding over a closed path surrounding the defect. Motivated by this, we demonstrate, in this Letter, that the single carbon defects at A and B sublattices of graphene can be regarded as pseudospin-mediated atomic-scale vortices with angular momenta $l = +2$ and $-2$, respectively. The quantum interferences measurements of the interacting vortices indicate that the vortices cancel each other, resulting in zero total angular momentum, in the $|A| = |B|$ case, and they show aggregate chirality and angular momenta similar to a single vortex of the majority in the $|A| \neq |B|$ case, where $|A|$ ($|B|$) is the number of vortices with angular momenta $l = +2$ ($l = -2$).**

Vortex is familiar to us because of the classical version, such as water vortex and hurricane vortex, and it is now recognized as a universal and significant phenomenon in various fields such as fluid physics, nonlinear optics, Bose–Einstein condensates, and condensed matter physics [1-10]. Recently, many attempts have been devoted to realize magnetic vortices based on the electronic spin in magnetic materials. However, reducing the size of a magnetic vortex to the atomic limit is difficult because changing the directions of spin at such a length scale will dramatically increase the energy of the system [11-18]. For electrons in graphene, besides the real electronic spin, there is an additional degree of freedom, *i.e.*, sublattice pseudospin, that arises from the unique bipartite honeycomb lattice structure of graphene [19-23]. It is easy to change the direction of the pseudospin at atomic scale without cost energy of the system. Therefore, it is possible to realize pseudospin-mediated atomic-scale vortex in graphene.

Very recently, it was demonstrated that the intervalley scattering induced by an atomic defect in graphene can lead to a rotation of pseudospin of the chiral quasiparticles [24,25]. Motivated by this, here we demonstrate that a single carbon defect at the A (B) sublattice of graphene can be regarded as a pseudospin-mediated atomic-scale vortex with angular momenta $l$ = +2 ($l$ = -2). The quantum interferences of the atomic-scale vortices with the same or opposite angular momenta are systematically studied both in experiment and theory. The interacting vortices cancel each other, resulting in zero total angular momentum, in the $|A|$ = $|B|$ case, where $|A|$ ($|B|$) is the number of vortices generated by the defects at the A (B) sublattice. In the $|A| \neq |B|$ case, the interacting vortices show aggregate chirality and angular momenta similar to a single vortex of the majority.

Generally, a topological vortex can be described as the phase winding of the wavefunction $\psi(\boldsymbol{r}) = f(\boldsymbol{r})e^{il\varphi_r}$ surrounding a phase singularity with zero wavefunction $f(\boldsymbol{r}_0) = 0$. Here $\varphi_r$ represents the azimuthal angle and $l$ is the angular momentum, representing the times of wavefunction rotate (winding number) when $\varphi_r$ undergoes a closed trajectory. Figures 1(a) and 1(b) show vectors of the wavefunction for the $l$ = +2 and -2 vortices, respectively. The + (-) donates clockwise (counterclockwise) circulation, *i.e.*, the chirality of the vortices. Since the topological and chiral features of vortex cannot be directly imaged, a proposal to capture the features of vortex via the interference has been widely adopted [26]. By introducing a plane wave that propagates downward, there are $N = l = \pm 2$ additional wavefronts in the interference patterns, as shown in Figs. 1(c) and 1(d). The number of additional wavefronts $|N| = 2$ indicates the angular momentum of the vortex, and the appearance of the additional wavefronts behind or ahead the vortex can directly reflect the chirality, + or -, of the vortex. Therefore, the topological and chiral features of the vortex are measurable via the interference patterns [24-29]. In graphene monolayer, the elastic intervalley scattering induced by the single carbon defect rotates the pseudospin by $\pm 2\theta_q$ (see Figs. 1(e) and 1(f), here $\theta_q$ is the incident angle of electrons with momentum $q$) [24,31-34] and an accumulation of the phase shift over a closed path enclosing the single carbon defect is $\pm \int_0^{2\pi} 2d\theta_q = \pm 4\pi$ [24]. Such an effect leads to $N = \pm 2$ additional wavefronts in the modulated charge densities because that each additional wavefront contributes $2\pi$ in the phase shift. Our low-energy continuum model calculations also reveal that the single carbon defect at the A and B sublattices (A-site and B-site defects) of graphene can generate $N = l = \pm 2$ additional wavefronts in the modulated charge densities for a selected direction of the intervalley scattering (Figs. S2 [30]). Therefore, the A-site or B-site defect in monolayer graphene should be regarded as a phase singularity that is responsible for the generation of the $l = +2$ or $l = -2$ atomic-scale vortex.

To explore the atomic-scale vortex nature of the single carbon defects, we carried out scanning tunneling microscopy (STM) measurements. At a given STM tip position, the charge densities are governed by the interference of the electronic waves between electrons in the tip pointing towards the single carbon defect and their reflection from the defect by coupling a phase shift [24,25]. Meanwhile, the STM tip can probe the local density of states (LDOS) of electrons with high spatial resolution [35,36]. Therefore, we can obtain the interference patterns between a tip-introduced plane electronic wave and a defect-induced atomic-scale vortex from the STM images. In our experiments, we carried out measurements of decoupled topmost graphene monolayer with a high density of single carbon defects on multilayer graphene, which was directly synthesized on Ni foils using a facile chemical vapor deposition (CVD) method [38-43] (see methods and Fig. S3 of the Supplemental Material [30]). Figures 1(g) and 1(h) show representative atomic STM images of the single-carbon defects at the A and B sublattices, respectively, acquired from a single-crystal monolayer graphene. Due to the Jahn-Teller distortion around the single carbon defect [44-47], the $C_2$ symmetry instead of the $C_{3v}$ symmetry induced by an adsorbed hydrogen is expected to be observed around the defects in the STM images [38,39]. Moreover, the distinctive topographic fingerprint of the triangular $\sqrt{3} \times \sqrt{3}$ R 30° interference patterns induced by the single carbon defect are clearly observed [38,39,48-50].

The fast Fourier transform (FFT) analysis of the STM images are shown in the inset of Figs. 1(i) and 1(j), respectively. The outer bright spots connected by yellow dashed hexagon are the reciprocal lattices of monolayer graphene, which reflect the same lattice orientations of the single-crystal monolayer graphene shown in Figs. 1(g) and 1(h). Therefore, the different orientations of the defect-induced tripod shapes (blue and green dotted outlines) in the STM images demonstrate that the two single carbon defects are at the A and B sublattices respectively (they exhibit the inversion symmetry with respect to the center of a C-C bond). At the center of the FFT images, a bright disk is observed, which is a significant feature of electronic properties in the monolayer graphene due to the forbidden intravalley backscattering [31-34]. The additional inner bright spots, at the corners of Brillouin zone connected by green dashed hexagon, are generated by the defect-induced intervalley scattering [31-34]. To explore the quantum interference of the intervalley scattering induced by the single carbon defect in monolayer graphene, we carried out the FFT-filtered analysis to obtain the modulation of charge densities due to the intervalley interference. Figures 1(i) and 1(j) show the FFT-filtered STM images along the directions of the intervalley scattering enclosed by the white circles. A clear signature of $N = +2$ ($N = -2$) additional wavefronts is observed for the A-site (B-site) defect, which is well consistent with our theoretical calculations (see Fig. S2 [30]). In our experiment, the same features are also observed in other directions of the defect-induced intervalley scattering in both the STM images and STS maps, and the result is quite robust under different experimental conditions, such as the applied bias voltages, tunneling currents, and rotation of the scanning angles (Figs. S4-S6 [30]). Therefore, the A-site and B-site defects in monolayer graphene can be regarded as the $l = +2$ and -2 atomic-scale vortices, respectively.

Now we begin to explore the quantum interferences between the pseudospin-mediated atomic-scale vortices in monolayer graphene. Figure 2 summarizes the interferences of two atomic-scale vortices with the same chirality (the same angular momenta), which are realized by two individual single carbon defects at the A sublattice (labeled as A-A defects). Figures 2(a) and 2(b) show STM images of the A-A defects with the distances between the two defects as $d = 9.3$ nm and $d = 2.7$ nm, respectively. Figures 2(c) and 2(d) show the corresponding FFT-filtered STM images along the directions of the intervalley scattering enclosed by the white circles (the insets). It is interesting to find that the overall number of the additional wavefronts induced by the two A-site defects is still 2 rather than 4, regardless of their distance. Such a result, at first glance, is counterintuitive, but can be well understood with the picture of two interacting vortices. As shown in top panels of Figs. 2(e) and 2(f), the winding number of vectors over a closed path surrounding the two $l = +2$ vortices is still 2, which is the same as that of a single $l = +2$ vortex shown in Fig. 1(a). Therefore, the two interacting vortices behave as a new vortex with $l = +2$ and, consequently, generate $N = 2$ additional wavefronts (bottom panels of Figs. 2(e) and 2(f)). Even though the total number of the additional wavefronts induced by the two A-site defects is irrespective of their distance, the detailed features of the additional wavefronts are strongly affected by the distance. When the distance of the two A-site defects is relatively large, i.e., $d = 9.3$ nm (Fig. 2(a)), the $N = 2$ additional wavefronts are observed around each of the two defects, accompanied by the opposite $N = -2$ additional wavefronts between the defects (Fig. 2(c)). Such a feature is reproduced quite well in the quantum interferences between two vortices with a relative large distance, as shown in Fig. 2(e). In this case, the structure of each vortex only exhibits slight deformation and the angular momenta for the vortex are still $l = +2$. Therefore, the $N = 2$ additional wavefronts are expected to appear around each vortex. When the distance of the two A-site defects is reduced to $d = 2.7$ nm, the strong interference completely deforms the structure of the vortices and makes them merge into a new $l = 2$ vortex, thus resulting in the $N = +2$ additional wavefronts together, as shown in Figs. 2(d) and 2(f).

Figure 3 summarizes the interferences of two atomic-scale vortices with the opposite chirality, which are realized by two individual single carbon defects at different sublattices of graphene (labeled as A-B defects). Obviously, the interacting vortices cancel each other, resulting in zero total angular momentum. Such a result is expected because that the winding number of vectors over a closed path surrounding the two vortices becomes zero, as shown in Figs. 3(e) and 3(f). Similar to the result obtained from the A-A defects with a large distance, the $N = +2$ and -2 additional wavefronts can be observed around the A-site and B-site defects respectively when the distance between them is relatively large. Then, the structure of each vortex still exhibits the $l = +2$ ($l = -2$) angular momentum as an isolated vortex, as shown in Fig. 3(e). It is worth noting that there exists a slight phase flip in the Fourier-filtered wavefront dislocations in the experimental date. There are many possible origins, such as the localized strain, the subtle interlayer coupling, the impurities intercalated between the graphene layers, and the slight wrinkle or ripple structure of the graphene surface. However, these effects do not destroy the topological features of monolayer graphene. Here we only concentrate on the numbers of additional wavefronts in the vicinity of the single carbon defects, which are reproduced quite well with the picture based on the quantum interferences between the vortices. More strict calculations based on the low-energy continuum model in the framework of a $T$-matrix approach also give the same result (see Supplemental Material [30] for details).

The above results demonstrate that the quantum interferences between a $l = +2$ vortex and a $l = -2$ vortex (antivortex) result in zero total angular momentum, and the two interacting vortices with the same chirality (e.g. $l = +2$) exhibit aggregate chirality and angular momenta similar to a single $l = +2$ vortex. Furthermore, similar result can be extended to the quantum interferences among multiple pseudospin-mediated atomic-scale vortices in monolayer graphene. For example, when there are three single carbon defects (vortices), either all the three have the same chirality or two of them have the same chirality and the third one has the opposite chirality, the resulting total number of additional wavefronts is always 2 (see Fig. 4 for details of experimental results and analysis). This can be easily understood because that the winding number of vectors over a closed path surrounding the three vortices is always 2, as shown in Figs. 4(e) and 4(f). More generally, we can obtain that the interacting vortices cancel each other, resulting in zero total angular momentum, in the $|A| = |B|$ case, and they show aggregate chirality and angular momenta similar to a single vortex of the majority in the $|A| \neq |B|$ case, where $|A|$ ($|B|$) is the number of vortices with angular momenta $l = +2$ ($l = -2$).

In summary, we demonstrate that the individual single carbon defect at the A or B sublattice of monolayer graphene can be regarded as a pseudospin-mediated atomic-scale vortex with the angular momenta of $l = +2$ or $-2$, respectively. The quantum interferences of the pseudospin-mediated vortices are systematically studied. Our result highlights the way to tailor the atomic-scale vortices in systems with pseudospin degree of freedom.

**Acknowledgements**

This work was supported by the National Natural Science Foundation of China (Grant Nos. 11974050, 11674029). L.H. also acknowledges support from the National Program for Support of Top-notch Young Professionals, support from "the Fundamental Research Funds for the Central Universities", and support from "Chang Jiang Scholars Program".

**Figures**

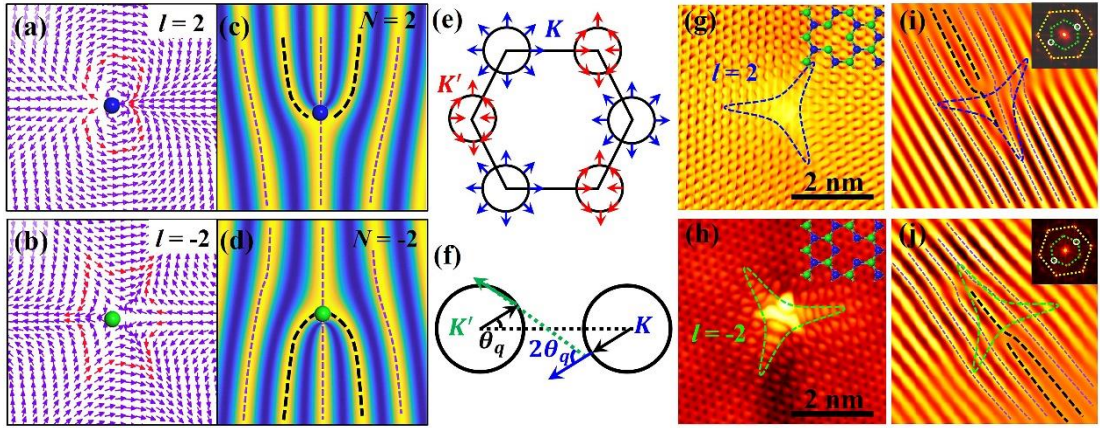

**Figure 1.** The wavefront dislocations induced by single carbon defect in monolayer graphene. **(a,b)** The vectors of wavefunction for $l = \pm 2$ vortices. The center of vortices is marked by blue or green dots. **(c,d)** Interference patterns between a vortex and a plane wave that propagates downward. The additional wavefronts are marked by black dashed lines. **(e)** Pseudospin textures along the Fermi surfaces in monolayer graphene. **(f)** Quasiparticles scattering from a given valley K′ (K) to a nearest valley K (K′) in graphene. The pseudospin rotates by $\pm 2\theta_q$. **(g,h)** The topography STM images of an individual single carbon defect in monolayer graphene ($V_b = 200\ mV$, $I = 300\ pA$). The triangular interference patterns induced by single carbon defect are marked by blue and green dotted outlines, which are related to the single carbon defect at the A and B sublattices, respectively. The atomic structures are given in the insets. **(i,j)** FFT-filtered images of (g,h) along the direction indicated by white circles. The black dashed lines correspond to $N = \pm 2$ additional wavefronts. Insets: FFT of the STM images in (g) and (h), respectively. The outer hexangular spots (corners of the yellow dotted line) and inner bright spots (corners of the green dotted line) correspond to the reciprocal lattice of graphene and the interference of the intervalley scattering, respectively.

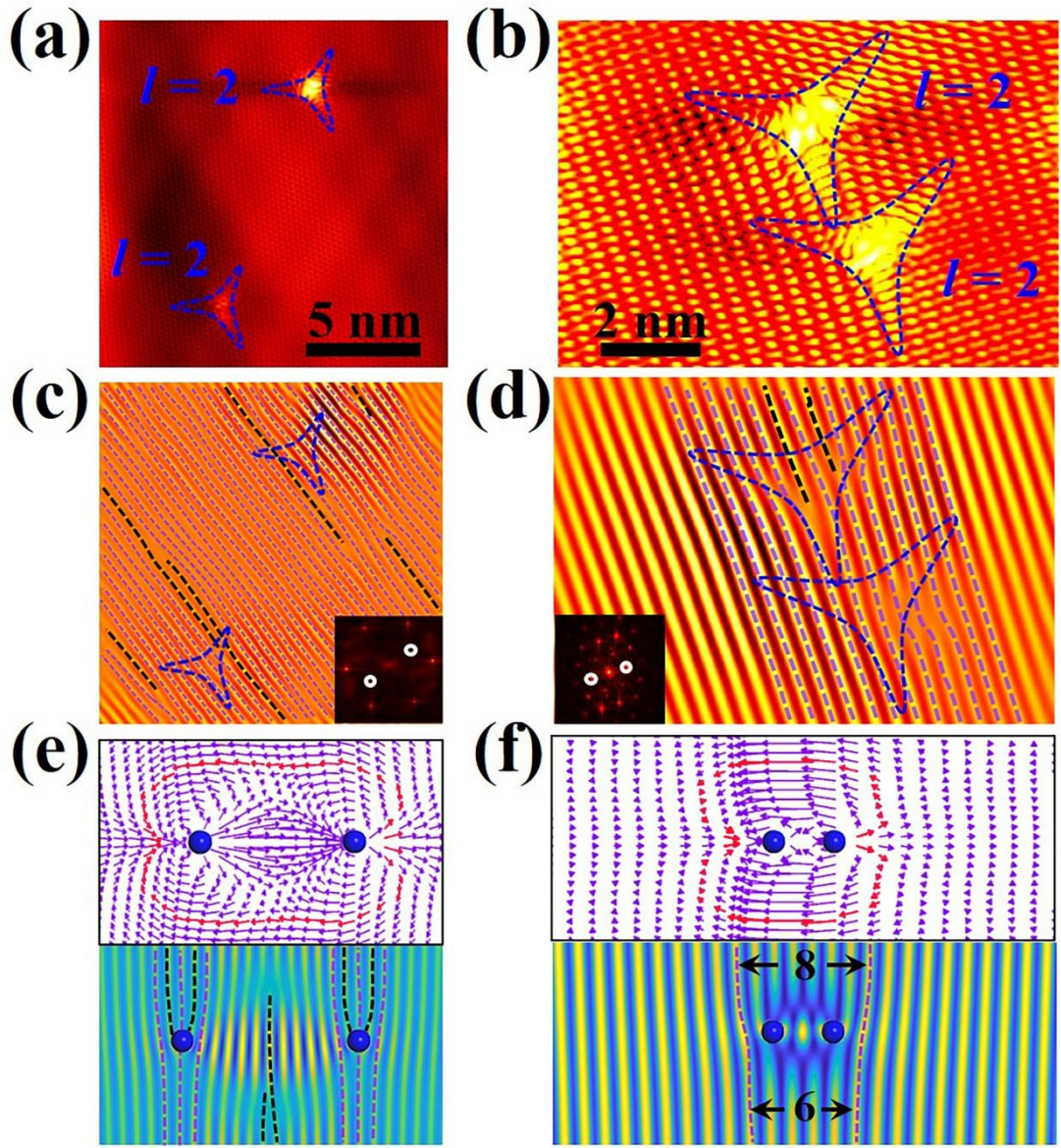

**Figure 2.** The interference of vortices and wavefront dislocations induced by A-A defects in monolayer graphene. **(a,b)** Typical STM images of the A-A defects in monolayer graphene with the separated distances of (a) 9.3 nm and (b) 2.7 nm, respectively. The dotted tripod shapes are added manually to indicate the orientation and position of the defects. **(c,d)** FFT-filtered images of (a) and (b) along the marked direction of the intervalley scattering. The additional wavefronts are marked by black dashed lines. Insets: the filters applied in the Fourier space. **(e,f)** Up panels: the structures of interference between two $l = +2$ vortices with different separations. Bottom panels: The interference patterns between two vortices and a plane wave propagating downward. The additional wavefronts are highlight in the figures.

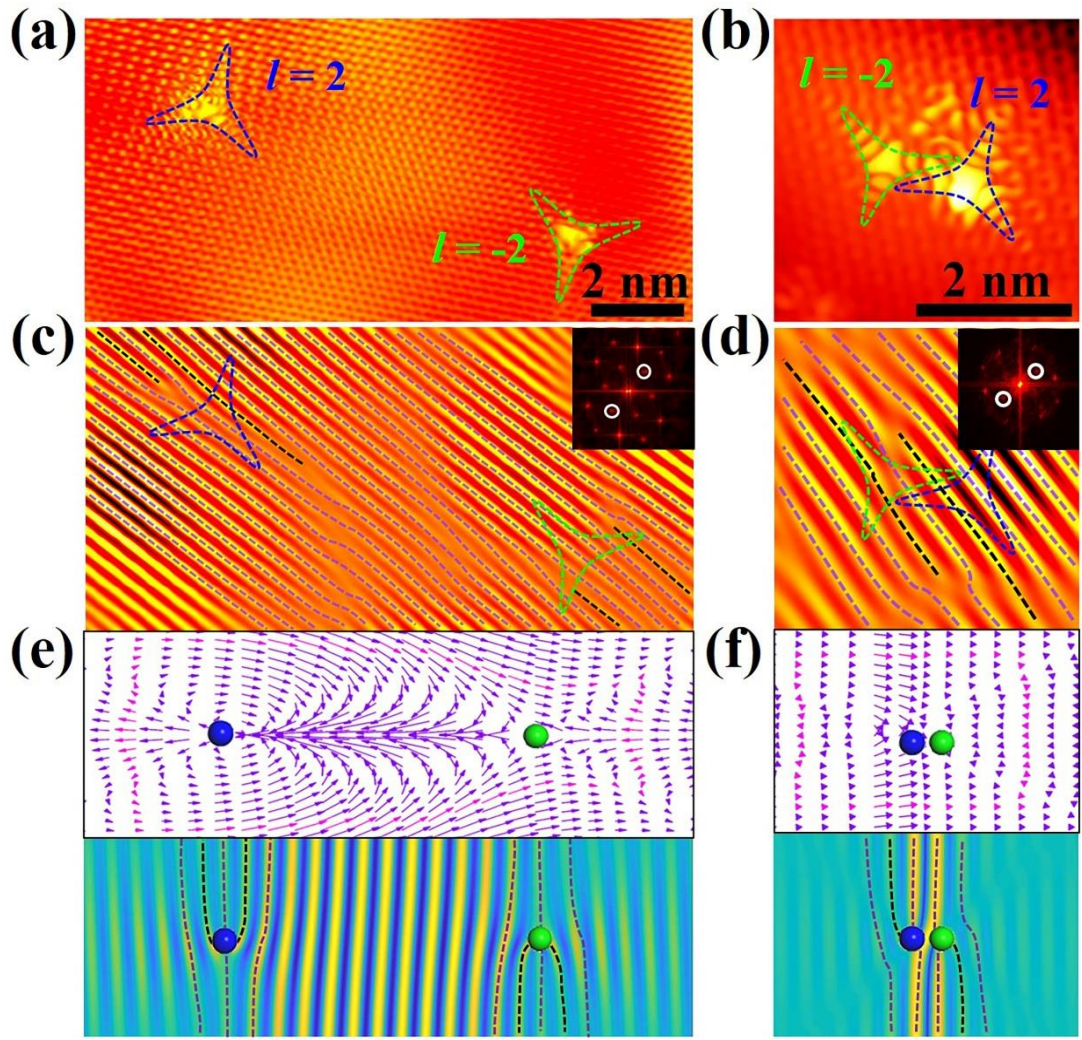

**Figure 3.** The interference of vortices and wavefront dislocations induced by A-B defects in monolayer graphene. **(a,b)** Typical STM images of the A-B defects in monolayer graphene with the separated distances of (a) 8.8 nm and (b) 1.0 nm, respectively. The dotted tripod shapes are added manually to indicate the orientation and position of the defects. **(c,d)** FFT-filtered images of (a) and (b) along the marked direction of the intervalley scattering. Insets: the filters applied in the Fourier space. **(e,f)** Up panels: the structures of interference between a $l = +2$ and a $l = -2$ vortices with different separations. Bottom panels: The interference patterns between two vortices and a plane wave propagating downward.

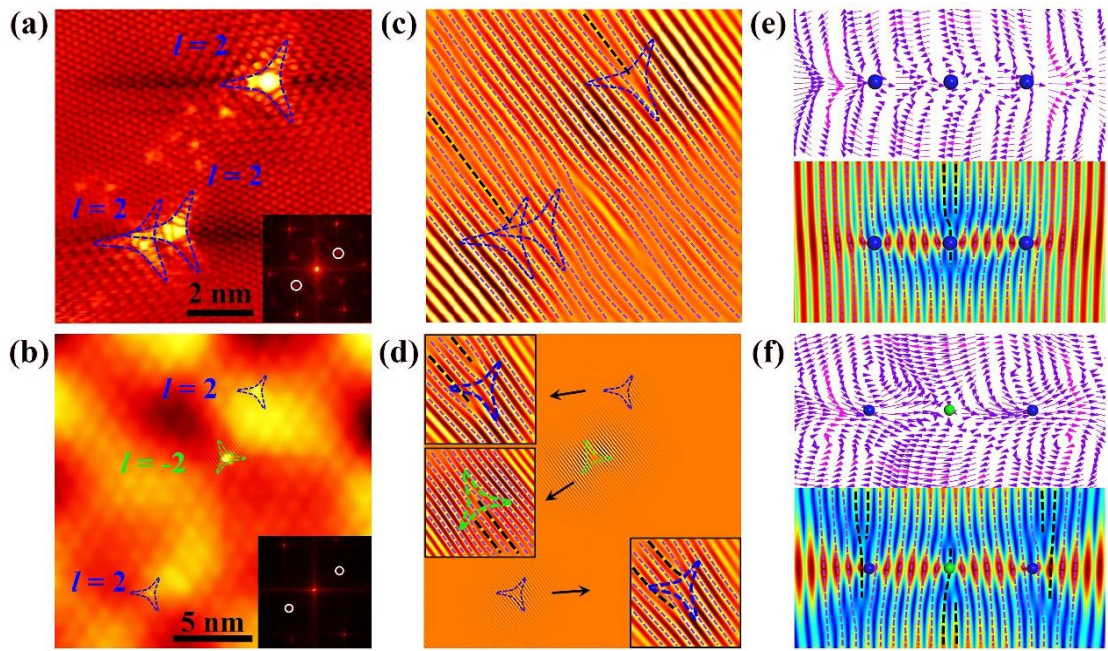

**Figure 4.** The interference of vortices and wavefront dislocations induced by A-A-A and A-B-A defects in monolayer graphene. **(a,b)** Typical STM images of A-A-A and A-B-A defects in monolayer graphene, respectively. **(c,d)** FFT-filtered images of (a) and (b) along the marked direction of the intervalley scattering. **(e,f)** Up panels: the structures of interference among panel (e) three $l = +2$ vortices and panel (f) two $l = +2$ vortices and one $l = -2$ vortices, respectively. Bottom panels: The interference patterns between three vortices and a plane wave propagating downward.